# THE EXTRAGALACTIC DISTANCE SCALE

SIDNEY VAN DEN BERGH

Dominion Astrophysical Observatory

National Research Council of Canada

5071 West Saanich Road

Victoria, British Columbia

V8X 4M6, Canada

## ABSTRACT

Cepheid variables are used to derive a Virgo cluster distance of 16.0 ± 1.5 Mpc. In conjunction with the Coma velocity and the well-established Coma/Virgo distance ratio, this yields a Hubble parameter $H_o$ = 81 ± 8 km $s^{-1}$ $Mpc^{-1}$. By combining this value with an age of the Universe $\gtrsim$ 16.8 ± 2.1 Gyr, that is derived from the metal-poor globular cluster M92, one obtains f (Ω, Λ) $\gtrsim$ 1.39 ± 0.22. This value is only marginally consistent with an Einstein-de Sitter universe with Ω = 0 and Λ = 0, which has f = 1. An Einstein-de Sitter universe with Ω = 1 and Λ = 0, for which f = ⅔ , appears to be excluded at the 3σ level. It is shown that some recent small values of $H_o$ resulted from the large intrinsic dispersion in the linear diameters of galaxies, and from the fact that well-observed supernovae of Type Ia exhibit a luminosity range of ~30 at maximum light.

## 1. INTRODUCTION

Seventy-six years ago Curtis (1921) and Shapley(1921) met in this auditorium to present their differing views on the nature of the nebulae. Shapley believed that our Milky Way galaxy was a vast continent surrounded by small island nebulae. On the other hand, Curtis advocated the view that spiral nebulae were very distant objects with dimensions similar to that of our own Milky Way System. This argument was decisively settled when Hubble (1925) discovered Cepheid variable stars in the nebulae M31 and M33. In the words of Sandage (1961): "No one knew [what galaxies were] before 1900. Few people knew in 1920. All astronomers knew after 1924".

The astronomical community is presently engaged in a similar high-stakes debate about the extragalactic distance scale. If the Hubble parameter $H_o \approx 50$ km $s^{-1}$ $Mpc^{-1}$, then we might live in an elegant and simple Einstein-de Sitter universe. However, if $H_o \approx 80$ km $s^{-1}$ $Mpc^{-1}$, then we exist in a more complex (and perhaps more interesting) world.

## 2. AGE CONSTRAINTS ON THE HUBBLE PARAMETER

The age of the Universe cannot be smaller than that of the oldest stars. The ages of the oldest star clusters therefore place severe constraints on the expansion age of the Universe. Recently Bolte & Hogan (1995) have fitted the main sequence of the very old metal-poor Galactic globular cluster M92 to that of nearby sub-dwarfs with similar metallicities, for which distances can be determined by trigonometric parallaxes. Stellar evolutionary tracks with an age of 15.8 ± 2.1 Gyr give an excellent fit to the color-magnitude diagram of M92 that has been calibrated by nearby sub-dwarfs. Recently, Chaboyer et al. (1996) have concluded that there is a 95% probability that the ages of the oldest globular clusters are > 21.1 Gyr. On the assumption that it requires $\gtrsim$ 1 Gyr for matter formed in the hot "Big Bang" to condense into stars, it follows, from the work of Bolte & Hogan, that the age of the Universe must be $t_o \gtrsim 16.8 \pm 2.1$ Gyr. This conclusion can only be avoided if current stellar evolutionary models are grossly in error.

The age of an expanding universe may be expressed in the form

$$t_o = f\,(\Omega, \Lambda)\, H_o^{-1}\,, \tag{1}$$

in which $f\,(\Omega, \Lambda) = ⅔$ for a "standard" Einstein-de Sitter universe with $\Omega = 1$, $\Lambda = 0$ and $f\,(\Omega, \Lambda) = 1$ if $\Omega = \Lambda = 0$. It follows from Eqn. (1) that

$$H_o\ (\mathrm{km\ s^{-1}\ Mpc^{-1}}) = 978\ f\ t_o^{-1}\ (\mathrm{Gyr}). \tag{2}$$

Substitution of $t_o \gtrsim 16.8 \pm 2.1$ Gyr into Eqn. (2) yields $H_o \lesssim 39 \pm 5$ km $s^{-1}$ $Mpc^{-1}$ for the standard Einstein-de Sitter universe with $\Omega = 1$, $\Lambda = 0$ and $H_o \lesssim 58 \pm 7$ km $s^{-1}$ $Mpc^{-1}$ for a universe with $\Omega = 0$ and $\Lambda = 0$. These results show that observations of the extragalactic distance scale can place severe constraints on permissible models of the Universe.

## 3. THE DISTANCE SCALE

Galaxies are exceedingly distant. Standard Candles in them are therefore dim and difficult to observe. As a consequence of this, determination of the extragalactic distance scale is a very difficult enterprise that challenges the utmost limits of our observational capabilities. "There, we measure shadows, and search among ghostly errors of measurement for landmarks that are scarcely more substantial" (Hubble 1936).

We now know that the first attempt by Hubble & Humason (1931) to determine $H_o$ was in error by a factor of 7 - 10. Using the brightest "stars" in distant galaxies as standard candles, Hubble & Humason found $H_o = 559$ km $s^{-1}$ $Mpc^{-1}$ , with what they estimated to be an uncertainty of about 10%. Subsequently Baade (1954) showed that confusion between classical Cepheids of Population I and W Virginis variables of Population II had resulted in an under-estimate of the distance scale by a factor of two, so that $H_o$ ~280 km $s^{-1}$ $Mpc^{-1}$. Using galaxy luminosity functions, Behr (1951) showed that the distance moduli of galaxies adopted by Hubble (and by Baade) should be increased by $\Delta(m\text{-}M) = 1.7 \pm 1.1$ mag, corresponding to a factor of ~2 in distance. Finally Sandage (1958) was able to show that many of the objects which Hubble had regarded as brightest stars in distant spirals were, in fact, HII regions. Correcting for this error Sandage arrived at a Hubble parameter $H_o \approx 75$ km $s^{-1}$ $Mpc^{-1}$ "with a possible uncertainty of a factor or 2". There the matter rested for the next 38 years. Kennicutt et al. (1995) have recently published an instructive plot showing all values of $H_o$ published during the last two decades. This figure shows that the vast majority of recent determinations fall in the interval $50 \lesssim H_o$ (km $s^{-1}$ $Mpc^{-1}$) $\lesssim 100$, with only a smattering of values (mostly derived by Sandage and his collaborators) falling below this range. For recent reviews of the distance scale problem, the reader is referred to Fukugita, Hogan & Peebles (1993), Jacoby et al. (1992) and van den Bergh (1992, 1994). A sampling of modern determinations of $H_o$ , based on a variety of different techniques, is shown in Table 1.

Inspection of Table 1 shows that the smallest values of $H_o$ have mostly been based on the use of supernovae of Type Ia (SNe Ia) as distance indicators. Perhaps, it is therefore not surprising that advocates of small values of $H_o$ have placed considerable emphasis on the use of SNe Ia as standard candles. This question will be discussed in more detail in § 4 of the present paper. In § 5 it will be shown that galaxy diameters are not "standard meter sticks". Values of $H_o$ derived from galaxy diameters are therefore of questionable significance.

$H_o$ estimates based on the Sunyaev - Zel'dovich effect have recently been discussed in great detail by Inagaki, Suginohara & Suto (1995). These authors conclude that uncertainties in the determination of temperature profiles of distant clusters can produce significant errors in $H_o$ values determined by using the S-Z effect. Furthermore, clusters are selected by surface brightness. Such selection introduces a bias in favor of clusters that are elongated along the line of sight. This will result in a bias towards low values of $H_o$ (Fukugita 1995). This bias is avoided for the nearby Coma cluster for which the S-Z method gives $H_o = 74^{+29}_{-24}$ km $s^{-1}$ $Mpc^{-1}$ (Herbig et al. 1995, Meyers et al. 1995).

Observation of gravitational lenses holds great promise for future determinations of the Hubble parameter. The present uncertainty in the difference in light-travel time $\Delta t$ for the lens 0957 + 561 should be removed by additional observations being undertaken this year. A more fundamental difficulty is that the value of $\chi^2$ for the model of 0957 + 561 recently given by Grogin & Narayan (1995) is large. It is therefore not yet clear how well their mass model captures the actual mass distribution in the lensing object.

For the remaining estimates of $H_o$ listed in Table 1 there is a clear-cut dichotomy between SNe Ia, which appear to give values of $H_o$ in the range 50 - 60 km $s^{-1}$ $Mpc^{-1}$ , and the other techniques that yield $H_o \gtrsim 70$ km $s^{-1}$ $Mpc^{-1}$ .

## 4. ARE SNe Ia GOOD STANDARD CANDLES?

Supernovae of Type I were first used as standard candles by Kowal (1968). A generally optimistic assessment of the value of SNe I as standard candles was given in the workshop on Supernovae as Distance Indicators (Bartel 1985). For example, Cadonau, Sandage & Tammann (1985) write: "Individual SNe I generally show no systematic deviations from these templet light curves, and occasional deviations are explained as photometric errors, which can be quite severe in the case of SNe. The peak luminosity of absorption-free SNe I is also uniform with an intrinsic rms scatter of < 0.3 mag". Shortly afterwards, it was, however, discovered (e.g. Harkness & Wheeler 1990) that there are at least two physically distinct types of SN I. Objects of Type Ia (SNe Ia) are now believed to have progenitors that belong to an old stellar population, whereas SNe Ib/c are thought to be produced by massive young progenitors. Perhaps the most stunning blow to the hypothesis that SNe Ia are good standard candles was provided by the discovery of the sub-luminous object SN1991bg. This supernova was observed to have B(max) = 14.75 (Filippenko et al. 1992, Leibundgut et al. 1993), which is ~2.5 mag (10x) fainter than SN1957B, which also occurred in the same (dust free!) elliptical galaxy NGC 4374 (= M84). Furthermore, the same year 1991 brought the discovery of the super-luminous SN Ia 1991T (Phillips et al. 1992). The discovery of both a super-luminous and a sub-luminous SN Ia made 1991 an annus horribilis for the hypothesis that SNe Ia are good standard candles.

Two "epicycles" have been proposed in attempts to save the hypothesis that SNe Ia are useful standard candles: (1) Branch, Fisher & Nugent (1993) have suggested that these objects may be segregated into "normal" SNe Ia that are good standard candles and spectroscopically peculiar ones that are not. However, Maza et al. (1994) show that the spectroscopically normal SNe Ia 1992bc and 1992bo differed in $M_B$ (max) by 0.8 ± 0.2 mag, i.e. by a factor of two in luminosity. (2) Phillips (1993) has proposed that the maximum luminosities of SNe Ia are closely correlated with their rates of decline and that SNe Ia could therefore still be used as standard candles after $M_B$ (max) has been corrected for the rate of decline in blue light $\Delta m_{15}$. Phillips found that

$$M_B\,(\max) = a + b\,\Delta m_{15}\,, \qquad (3)$$

with b = -2.70. From more extensive data, Hamuy et al. (1995) have recently found a much smaller value b = -1.62. Faith in the usefulness of Eqn. (3) is further undermined by the fact that both SN 1885 = S And (de Vaucouleurs & Corwin 1985) and SN 1994D (Patat et al. 1996) exhibit large deviations from the mean maximum magnitude versus rate-of-decline relation. This suggests that departures from this relation (if it exists at all) may be quite large. It is not yet clear whether multi-parameter light curve models (Riess, Press & Kirshner 1995) will be able to fit strongly deviant objects such as S Andromedae.

Recently van den Bergh (1996) has determined the values of $M_B$(max) for all supernovae of Type Ia for which Cepheid distances are known. These data are collected in Table 2. This table shows that well-observed supernovae of Type Ia exhibit a range of ~30 in luminosity at maximum light. This result clearly places the usefulness of SNe Ia as calibrators of the extragalactic distance scale in doubt.

The data in Table 2 appear to confirm the suspicion that SNe Ia in late-type galaxies are, on average, more luminous that those in E and S0 galaxies (or in the nuclear bulges of spirals). However, a striking exception to this rule is provided by SN 1994D, which recently occurred in the Virgo cluster S0 galaxy NGC 4526 (Patat et al. 1996). For this object $B_o$(max) = 11.58 ± 0.08. With a Virgo distance modulus $(m\text{-}M)_o$ = 31.02 ± 0.2 (van den Bergh 1995a) this yields $M_B$(max) = -19.43 ± 0.22. Possibly, this high luminosity is due to the fact that SN 1994D had a relatively young progenitor that was associated with the prominent dust lane in NGC 4526.

## 5. GALAXY DIAMETERS AS STANDARD METER STICKS

Sandage (1993a) has used the assumption that supergiant spirals of type Sc I have constant linear diameters to derive $H_o$ = 43 ± 11 km $s^{-1}$ $Mpc^{-1}$. There are, however, a number of reasons for questioning the legitimacy of this technique for determining extragalactic distances:

### 5.1 Images of galaxies

Van den Bergh (1992) showed KPNO 4-m telescope images, printed to the same linear scale, of the Sc I galaxies NGC 309 in Cetus and M100 in Virgo. These images show that NGC 309 is two to three times as large as M100. This clearly shows that Sc I galaxies exhibit a considerable range in diameters.

### 5.2 Distance to M100

By assuming that M100 has the same diameter as the nearby Sc I galaxy M101 (in which Cepheids have been observed), Sandage (1993a) derived a distance of 27.7 Mpc to M100. However, recent observations of Cepheids in this galaxy with the Hubble Space Telescope (Farrarese et al. 1996) yield a distance of only 15.8 Mpc.

### 5.3 Comparison with M31 and M33

A comparison of the diameters of M31 and M33 with those of galaxies having similar classification types (van den Bergh, Pierce & Tully 1990) in the Ursa Major / Virgo clusters (Tully 1988) yields (van den Bergh 1992) distances of $17.1^{+3.1}_{-2.3}$ Mpc from M31 (D = 725 kpc assumed) and $10.6^{+3.2}_{-2.0}$ Mpc from M33 (D = 795 kpc assumed). These results show that (1) diameters of individual galaxies exhibit too large a range to make them useful as precision distance indicators, and (2) comparison of M31 with distant galaxies of type Sb I - II gives much larger distances than does a comparison of M33 with distant galaxies of type Sc II-III. This result undermines the conclusions of Sandage (1993b), which was only based on comparisons of M31 with distant galaxies of similar type.

On the basis of the results presented above, it is clear that galaxy diameters exhibit too large a dispersion to make them suitable for precision determinations of the extragalactic distance scale.

## 6. THE DISTANCE SCALE FROM CEPHEIDS

The first great distance-scale debate was decisively resolved when Hubble (1925) discovered Cepheids in nearby galaxies. The availability of the Hubble Space Telescope (HST) now makes it possible to observe classical Cepheids in distant galaxies. Using such Cepheid distances the resolution of the second great distance-scale controversy now appears to be within reach.

Table 3 lists Cepheid distances to five spiral galaxies in the Virgo region. The first four objects in this table yield a consistent distance, whereas that of NGC 4639 gives a distance that is 9 Mpc larger. The Tully-Fisher distance of 22.3 ± 2.2 Mpc (Yasuda, Fukugita & Okamura 1996) also indicates that NGC 4639 is a background object. The first four spirals in Table 1 yield a formal weighted true Virgo distance modulus $(m\text{-}M)_o$ = 31.02 ± 0.08. To this error should be added a 0.1 mag systematic uncertainty resulting from possible errors in the zero-point of HST photometry and an uncertainty of ~0.1 mag in the distance modulus of the Large Magellanic Cloud, relative to which the Virgo distances were determined[1].

---

[1] The RR Lyrae magnitudes in LMC clusters (Walker 1952) might indicate that the distance to the Large Cloud has been over-estimated. If this is indeed the

case, then the value of $H_o$ derived below needs to be increased.

---

In the subsequent discussion, it will be assumed that the true distance modulus of the Virgo cluster is $(m\text{-}M)_o = 31.02 \pm 0.2$, corresponding to a distance of 16.0 ± 1.5 Mpc. Recently, Whitmore et al. (1995) have found $(m\text{-}M)_o = 31.12 \pm 0.26$ by comparing the luminosity function of globular clusters in the Virgo giant elliptical M87 to those of globular clusters in M31 and the Galaxy. This result appears to show that spirals listed in Tale 1 lie at the same distance as the core of the Virgo cluster. The peculiar motion of the Virgo cluster relative to the Hubble flow, and the infall velocity of the Local Group into the Virgo cluster, are not yet well-determined. It is therefore safest to derive the Hubble parameter from the Coma/Virgo distance ratio and the Coma cluster velocity relative to the microwave background. From 12 concordant determinations, van den Bergh (1992) found $\Delta(m\text{-}M)_o = 3.71 \pm 0.05$ for the difference between the Virgo and Coma distance moduli. With $(m\text{-}M)_o$ (Virgo) = 31.02 ± 0.2 and $\Delta(m\text{-}M)_o = 3.71 \pm 0.05$, one obtains $(m\text{-}M)_o$ (Coma) = 34.73 ± 0.21, corresponding to a distance D(Coma) = 88 ± 9 Mpc. Durret et al. (1996) find a mean redshift $< V > = 6901 \pm 72$ km $s^{-1}$ for the Coma cluster. With a correction of +258 ± 10 km $s^{-1}$ to place Coma in the cosmic microwave background frame, this yields a true velocity V(Coma) = 7159 ± 73 km $s^{-1}$. From these values one obtains $H_o$ = V(Coma) / D(Coma) = 81 ± 8 km $s^{-1}$ $Mpc^{-1}$. Substitution of this value and $t_o \gtrsim 16.8 \pm 2.1$ Gyr into Eqn. (2)

yields f $(\Omega, \Lambda) \gtrsim 1.39 \pm 0.22$, which is only marginally compatible with the f = 1.0 value for an Einstein-de Sitter universe with $\Omega = 0$ and $\Lambda = 0$. An Einstein-de Sitter model with $\Omega = 1$ and $\Lambda = 0$, for which $f = \frac{2}{3}$, appears to be ruled out at the $3\sigma$ level by Cepheid distances.

I am deeply indebted to Dr. Abi Saha for permission to quote his distances to spirals in the Virgo region in advance of publication, and to Dr. Brad Schaefer for discussions about the distances to individual supernovae of Type Ia.

TABLE 1

SAMPLE OF RECENT DETERMINATIONS OF $H_o$

| $H_o$ (km $s^{-1}$ $Mpc^{-1}$) | Technique | Reference |
|---|---|---|
| 86 ± 18 | PN in Virgo cluster | Mendéz et al. (1993) |
| 84 ± 8 | Fisher-Tully | Lu et al. (1994) |
| 81 ± 8 | Cepheids in 4 Virgo spirals | van den Bergh (1995a) |
| 80 ± 12 | SB fluctuations | Jacoby et al. (1992) |
| 78 ± 11 | Globulars in M87 | Whitmore et al. (1995) |
| 75 ± 8 | PN in Fornax cluster | McMillan et al. (1993) |
| 70 ± 13 | Novae in Virgo | Della Valle & Livio (1995) |
| 60 or 82[a] | Lens 0957 + 561 | Grogin & Narayan (1995) |
| 55 ± 17 | Sunyaev - Zel'dovich effect | Birkinshaw & Hughes (1994) |
| 55 to 60 | SNe Ia (theory) | van den Bergh (1995b) |
| 52 ± 9 | SNe Ia (1937C) | Saha et al. (1994) |
| 52 ± 8 | SNe Ia (1972E) | Saha et al. (1995) |
| 43 ± 11 | Galaxy diameters | Sandage (1993a) |

[a] The values $H_o$ = 60 km $s^{-1}$ $Mpc^{-1}$ and $H_o$ = 82 km $s^{-1}$ $Mpc^{-1}$ listed by Grogin & Narayan (1995) are for time differences of $\Delta t$ = 1.5 yr. and $\Delta t$ = 1.1 yr., respectively.

TABLE 2

SNe Ia WITH CEPHEID CALIBRATIONS

| SN | Galaxy | Type | $(m\text{-}M)_o$ | $\Delta m_{15}$ | $M_B$(max) |
|---|---|---|---|---|---|
| 1885 | M 31 | Sb | 24.3 ± 0.1 | 2.1 | -17.33: |
| 1937C | IC 4182 | Ir | 28.36 ± 0.11 | 1.25 | -19.53 ± 0.16 |
| 1960F | NGC 4496 | SBc | 31.1 ± 0.15 | 1.06 | -20.40 ± 0.79 |
| 1972E | NGC 5253 | Pec | 28.10 ± 0.12[a] | 1.1: | -19.55 ± 0.21 |
| 1981B | NGC 4536 | Sc | 31.10 ± 0.13 | 1.1 | -19.22 ± 0.28 |
| ... | Virgo[b] | ... | 31.02 ± 0.2 | < 1.1 > | -18.76 ± 0.24 |
| 1990N | NGC 4639 | Sb | 32.00 ± 0.23 | 1.15 | -19.43 ± 0.23 |
| [1991T | NGC 4527 | Sb | 31.05 ± 0.15[d] | 1.1 | -19.99 ± 0.21] |
| 1991bg[c] | NGC 4374 | E1 | 31.02 ± 0.2 | 1.9 | -16.27 ± 0.2 |
| 1994D | NGC 4526 | S0 | 31.02 ± 0.2 | 1.26 | -19.43 ± 0.22 |

[a] Visual distance modulus.

[b] Eight SNe in E and S0 galaxies that occurred before 1991.

[c] Spectroscopically peculiar.

[d] Van den Bergh (1996).

TABLE 3

DISTANCES TO SPIRALS IN THE VIRGO REGION

| Galaxy | $(m\text{-}M)_o$ | D(Mpc) | Reference |
|---|---|---|---|
| NGC 4321 | 31.00 ± 0.20 | 15.8 | Farrarese et al. (1996) |
| NGC 4496 | 31.10 ± 0.15 | 16.6 | Saha et al. (Saha 1995) |
| NGC 4536 | 31.05 ± 0.15 | 16.2 | Saha et al. (Saha 1995) |
| NGC 4571 | 30.91 ± 0.15 | 15.2 | Pierce et al. (1994) |
| NGC 4639 | 32.00 ± 0.23 | 25.1 | Sandage et al. (1996) |